\documentstyle[preprint,aps]{revtex}
\begin{document}
\draft
\vskip 0.5cm
\preprint{McGill/95-27}
\title{Lattice gas model for fragmentation:\\
	From Argon on Scandium to Gold on Gold}

\author{Subal Das Gupta \ and \ Jicai Pan}
\address{Department of Physics, McGill University \\
3600 University St., Montr\'{e}al, PQ, H3A 2T8 Canada}

%\date{}
\maketitle
\begin{abstract}
The recent fragmentation data for central collisions of Gold on Gold are
even qualitatively different from those for central collisions
of Argon on Scandium.  The latter can be fitted with a lattice gas
model calculation.  Effort is made to understand why the model fails for
Gold on Gold.  The calculation suggests that the large Coulomb interaction
which is operative for the larger system is responsible for this discrepancy.
This is demonstrated by mapping the lattice gas model to a molecular
dynamics calculation for disassembly.  This mapping is quite faithful
for Argon on Scandium but deviates strongly for Gold on Gold.  The molecular
dynamics calculation for disassembly reproduces the characteristics of the
fragmentation data for both Gold on Gold and Argon on Scandium.
\end{abstract}
%\pacs{25.70.-z, 24.60.-k,05.50.+q, 05.70.-a}
%\narrowtext

\newpage
\section{Introduction}

  Recently we proposed a lattice gas model which was used to calculate
mass distributions seen in heavy ion collisions at intermediate
energy \cite{pan1,pan2}.
There are several features of this model that are attractive.  The model
can be used to study liquid-gas phase transition in mean-field theory and
thus has links with Skyrme model studies of phase transitions.  But the
model can also be used to obtain cluster distributions and in this respect
has close ties with percolation model of fragmentation which has been used
\cite{bauer,campi}  with success in theoretical studies of heavy ion
collisions.
However, the lattice gas model has both kinetic energy and
interactions, thus the scope of the model goes beyond that of standard
percolation model.

The model was used to fit the data obtained in Michigan State University for
central collisions of Ar on Sc \cite{li1,li2}.  There are several features of
the
model that are quite general and are also seen in experiments.  First of all,
there is a region of beam energy where the yield of $Y(A)$ of fragments
as a function of the mass number $A$ of the fragment
obeys a power law first noted in pioneering experiments by the Purdue
group \cite{finn}.  This feature emerges in theoretical calculations also.
In the lattice gas model the input in a calculation is the temperature and on
general grounds the higher the beam energy, the higher
the temperature.  In the
lattice gas model there is a temperature at which a power law will emerge.
If in the vicinity of this critical temperature the mass distribution is
fitted by a power law whose exponent is denoted by $\tau$ then we expect
to see a minimum in the extracted value of $\tau$ at the critical temperature.
It is difficult to miss a minimum in $\tau$ in percolation type approach and
also in the lattice gas approach.  This minimum was seen in the Michigan
experiment and was found in the calculation of \cite{pan1,pan2} at the
experimental
beam energy.  Such a minimum in the value of $\tau$ has also been seen in a
recent experiment at Chalk River Nuclear Laboratories \cite{luc} and the fit
with a lattice gas model calculation
is quite pleasing \cite{pan3}.  Thus it seems that the lattice gas model
simulates
the fragmentation of nuclear matter reasonably well at least for medium mass
collisions.

Very recent data for Au on Au central collisions \cite{msu}
are at variance with these general expectations.
In the range of beam energy 35 MeV/nucleon to 100 MeV/nucleon no minimum in
$\tau$
was seen.  Since this is now a much bigger system and phase transition effects
should become more pronounced as we go to bigger systems, this absence of a
minimum in $\tau$ seems to cast serious doubts about the validity of the
model.  The other remarkable result is that at 35 MeV/nucleon of beam energy
the deduced value of $\tau$
is much below 2; it is 1.25.  Such a low value of $\tau$ will be difficult
to obtain in the lattice gas model even including shape effects as discussed in
\cite{phair}.  Thus if we apply the lattice gas model as proposed in
\cite{pan1}
and \cite{pan2} the calculation will not fit the data for the central
collisions
of Au on Au.

The present work started as an effort to understand this puzzle.  Between
Ar on Sc and Au on Au the main difference is not only the sizes
but also that in the latter case there is a huge Coulomb field whereas
presumably in the former case the Coulomb field is merely a minor perturbation.
Now the lattice gas model is a model of nearest neighbour interactions.
The Coulomb field is a long range force and thus is not amenable to lattice
gas type of approximation.  One way to investigate the effect of the Coulomb
field would be to try to add the effect of Coulomb force in the lattice
gas type of approach but we could find no obvious way of doing that.  We
had provided a prescription, based on simple physical reasoning, to decide
if two nucleons occupying neighbouring sites, form part of the same cluster
or not.  When in addition there is a long range force that distinguishes very
strongly between neutrons and protons, this criterion clearly needs to be
modified.  There is no simple way of knowing how the modification should be
made.
Hence to find what effects a strong Coulomb field may have on a lattice gas
model prediction, we need to try a somewhat convoluted approach.  We first
try to map the lattice gas model calculation to a molecular dynamics type
calculation, both first done without any Coulomb interaction.  If the
calculations match quite faithfully then we can study the effects of the
Coulomb interaction by adding that to the molecular dynamics calculation.
Hopefully we will find that for Ar on Sc the Coulomb interaction
does not severely change results and that the changes are very significant
for the case Au on Au.  We do not do an $ab~ initio$ molecular dynamics
calculation but only use it for disassembly from a thermally equilibrated
source.  The starting
point of the lattice gas model is that matter has equilibrated at some
temperature $T$.  There are $n$ nucleons and $N$ lattice sites ($N>n$).  Which
particular lattice sites are occupied are entirely dictated by statistical
mechanics.  Each cubic lattice has a size $1.0/\rho_0=6.25$ fm$^3$ and can,
at most, be occupied by a single nucleon.
This starting point of calculation will not be questioned in the present
work but we will use two different prescriptions for obtaining the yield
$Y(A)$ against $A$.  One of these two is our old prescription \cite{pan1,pan2}.
Nucleons occupying neighbouring sites will have attractive
interactions $-\epsilon$ and are considered to be
part of the same cluster provided the kinetic energy of relative
motion of these two nucleons does not overcome their binding.
This is enough information to deduce the yield $Y(A)$.  In the alternative
prescription that we will carry out here we fall back upon a more standard
many body calculation.
At the starting point when the nucleons have been initialised
at their lattice sites and have their initial momenta, we will switch to a
molecular dynamics calculation in which we will let the system
evolve according to
standard classical mechanics.  Nucleons which stay together after
arbitrarily long time are part of the same cluster.  After a sufficiently
long time the mass distribution is obtained.  This can now be compared with
lattice gas model predictions.  Obviously for this test
to be made with the lattice gas model we should choose for disassembly by
molecular dynamics an interaction which
suits the lattice gas model the best.  The potential should be deepest with
the value $-\epsilon$ when the two nucleons are $r_0=1.842$ fm apart and should
fall to 0 before $\sqrt{2}\times 1.842$ fm (the next nearest neighbour
interaction is zero in the lattice gas model ).  For distance less than
$1.842$ fm the potential should quickly become repulsive ( two nucleons can
not occupy the same site ).  If the two prescriptions match
for the yield $Y(A)$, then we have linked the lattice gas model prediction
for the yield $Y(A)$ against $A$
to a more well-known and better understood molecular dynamics approach.
Considering that the lattice gas model can be easily linked with percolation
model this in itself is quite interesting; we have provided a
connection between percolation model results and molecular dynamics which
seem to address totally different scenario to start out with.  Secondly, in
case the two
results match in the absence of a Coulomb field, we can, in the molecular
dynamics approach find out what a large Coulomb field, which can not
be incorporated in the lattice gas model, can do to the yield $Y(A)$ since the
Coulomb interaction is easily incorporated in molecular dynamics calculations.
We will give the necessary details of these two calculations in the next
section.

  We reiterate that our objective is not a molecular dynamics calculation
as such \cite{belkacem,lenk}; we use disassembly by molecular
dynamics with a nuclear force that produces results similar to those of lattice
gas model. This work is focussed towards one question only: we ask
why the lattice gas model which worked reasonably well for Ar on Sc
fails in the case of Au on Au and if we can relate this failure to the
large Coulomb field which is present in the second case.
We note also that the Coulomb interaction can be incorporated in
microcanonical models \cite{gross1,gross2}.

\section{Details of calculation}

Motivation and details of the lattice gas model are given in \cite{pan1} and
\cite{pan2}.
For completeness some of these details are provided in this section.
The calculations all require numerical simulation involving Monte-Carlo.
The starting point of all our calculations is this.  For a system of
$n$ nucleons we consider $N$ lattice sites where $N>n$.  $N$ is a parameter
chosen in \cite{pan1,pan2} by requiring the best fit.  The quantity $N/n=
\rho_0/\rho$ where $\rho_0$ is the normal density and $\rho$ is the so called
freeze-out density.  It will be seen in the following that in the lattice gas
calculation for fragments this freeze-out density is the actual density
at which all clusters are calculated.  For the molecular dynamics calculation
that we will perform $\rho$ is the density at which the initialisation is
done according to prescription of equilibrium statistical mechanics.  We then
let the system evolve in time and the cluster distributions are calculated
much later.  Thus strictly speaking $\rho$ is not a freeze-out density for
molecular dynamics calculation but merely defines the starting point
for time evolution.  However since classical evolution of a many particle
system is entirely deterministic, the initialisation does have in it all the
information of the asymptotic cluster distribution.  We will continue to
call $\rho$ the freeze-out density.

For initialisation, we assume that the nuclear part
of the interaction is simply $-\epsilon$ between nearest neighbours and
zero otherwise.
To begin a calculation we have to determine which of the sites
nucleons occupy and what their momenta are.  The two samplings can be done
independently of each other.  In a percolation model the $N$ sites would
be occupied with an occupation probablity $p=n/N$ by $n$ nucleons
in which each site has an equal
$a priori$ probability.  Because of interactions this is somewhat more
complicated in our case.  Let us assume we are handling the case where we will
take into account the Coulomb interaction explicitly in the initialisation.
Starting with all lattice sites empty, the first
nucleon ( a proton or a neutron as dictated by a Monte-Carlo decision )
is put at a site at random.  If this first nucleon is a neutron then the
probability of occupation of its nearest neighbours is proportional to
$\exp(\beta \epsilon)$
whereas all other sites have an occupation probability proportional to 1.
As usual, $\beta$ is the inverse of $kT$.  These probabilities
are now used to put in the second nucleon.  If the first nucleon was a proton
and the second one is a neutron then again the same probability of
occupation will be used.  But if the second nucleon is a proton also then
the above occupation probabilities are changed to proportional to
$\exp(\beta \epsilon-\beta u_c)$ for the nearest neighbours
and proportional to $\exp(-\beta u_c)$ for the other sites
where $u_c=e^2/r$, $r$ being the appropriate distance
between the two lattice sites.  It is obvious how to repeat this procedure
till the prescribed number of protons and neutrons are obtained.  It is also
obvious how to obtain the initial configuration when the Coulomb force is
not explicitly included.  In that case the prescription is identical with
what was used in \cite{pan1} and \cite{pan2}.

We do some initialisations where we take the Coulomb interaction explicitly
and  some initialisations when the Coulomb interaction is not taken into
account
separately.  The nuclear part is always characterised by a strength
$-\epsilon$ which is
the nearest neighbour type and has the same value irrespective of the isotopic
spin.  For a given nucleus the value of $\epsilon$ is lower for the case when
the Coulomb interaction is not explicitly added.  This is required by demanding
that the same binding energy is obtained in both the prescriptions.  In the
other case when the Coulomb interaction is explicitly included the nuclear
part of the interaction will lead to a larger binding which will be somewhat
compensated by the repulsive Coulomb part.  For the cases dealt with here
$\epsilon$ changes by roughly 1 MeV.

For disassembly by molecular dynamics we approximate the nuclear part of
the force by a well-known parametrisation \cite{stillinger}:
\begin{eqnarray}
v(r)&=&
\left\{
\begin{array}{ll}
A[B(r_0/r)^p-(r_0/r)^q]\exp[1/(r/r_0-a)], & \mbox{for $r/r_0<a$}\\
0,			& \mbox{for $r/r_0\ge a$}
\end{array}
\right.
\end{eqnarray}
Here $r_0$ is the distance between the centres of two adjacent lattices.  We
have
chosen $p=2$, $q=1$ and $a=1.3$.  The other constants $A$ and $B$ are chosen so
that
the potential acquires the prescribed value $-\epsilon$ at $r=r_0$.  With this
potential the interaction between two nucleons is zero when they are more than
$1.3r_0$ apart and the interaction begins to become strongly repulsive for $r$
significantly less than $r_0$ ; yet the potential is smooth enough that
accurate numerical solutions of time evolution of nucleons can be obtained.
The time evolution equations for each nucleon are, as usual, given by
$\partial{\bf p}_i / \partial t=-\sum _{j\ne i}{\bf \nabla}_i v(r_{ij})$
and $\partial{\bf r}_i/\partial t={\bf p}_i/m$.

The lattice gas predictions for cluster production can only be calculated
for the case where the Coulomb interaction is not explicitly included
but only through a lower value of $\epsilon$.  As mentioned already the
cluster distribution is calculated immediately after initialisation.
The lattice filling is done and the momenta are then generated from a
Monte-Carlo sampling of a Maxwell-Boltzmann distribution with a prescribed
temperature.  Nucleons in two neighbouring cells are considered to be part of
the same cluster if the kinetic energy of relative motion is not large enough
to
overcome the attractive interaction, i. e., $p_r^2/2\mu -\epsilon<0$.
Here $\mu$ is the reduced mass and is equal to $m/2$.
This definition is the simplest that one can provide and is physically
reasonable.
It should be emphasized that it is by no means a unique one.  The
prescription manages to reduce a many body problem of cluster production
into a sum of independent two body problems.  One can easily construct
scenarios where this prescription may underestimate the size of a cluster
and scenarios where this prescription may overestimate the size of a cluster.
It should be pointed out that this formula for bond formation has the same
structure as the one used in \cite{li2}.  Since each particle obeys the
Maxwell-Boltzmann distribution, the distribution of relative momentum between
two particles is also a Maxwell-Boltzmann, i.e.,
$P({\bf p}_r)=[1/(2\pi \mu kT)^{3/2}]\exp[-{\bf p}_r^2/2\mu kT]$.  We can then
write down a formula for the bonding probability which is temperature
dependent
\begin{eqnarray}
p=1-\frac{4\pi }{(2\pi \mu kT)^{3/2}}\int_{\sqrt{2\mu \epsilon}}^{\infty}
e^{-{\bf p}_r^2/2\mu kT}p_r^2dp_r
\end{eqnarray}
Switching to a variable $E=p_r^2/2\mu $ we get
\begin{eqnarray}
p=1-{\int_{\epsilon}^{\infty}e^{-E/kT}E^{1/2}dE \over
	\int _0^{\infty}e^{-E/kT}E^{1/2}dE}
\end{eqnarray}
which is identical with the formula of \cite{li2}.

For molecular dynamics calculation after initialisation we do the time
propagation long enough so that for cluster production the
asymptotic time has been reached.  Two
nucleons are part of the same cluster if the configuration distance between
them is less than 1.3$r_0$.  We stop the calculation after the original blob
of matter has expanded to 64 times it volume at initialisation.  For low
temperature this means doing the time evolution as long as $1000$ fm/c.
We use a time step of $0.1$ fm/c and update positions and momenta half a time
step apart ("leap frog" method).  The energy conservation in our calculation
is accurate to within 1 one percent.  The program conserves total momentum
identically.  Below we now consider specific cases.

\section{Results}

In Ref. \cite{pan1} we found that a freeze-out density $\rho=.39\rho_0$ gave
the
best fit with data.  Here we present data with this freeze-out density.  A few
calculations with a higher value of freeze-out density were also performed
but only to ascertain that the trends of the results are not strongly dependent
on the freeze-out density employed.

Fig. 1 shows the results of a $Y(A)$--$A$ plot of a lattice gas
calculation.  This is a repeat of the type of calculation done in
\cite{pan1,pan2}.  The value of $\epsilon$ used is 3.7 MeV.  This curve should
be compared with a molecular dynamics calculation shown in Fig. 2.  This
calculation uses the same $\epsilon$ and no explicit Coulomb interaction.  The
similarity between Figs. 1 and 2 is quite striking and leads us to conclude
that the simple prescription of cluster counting is very reasonable.  In Fig. 3
we have done a molecular dynamics calculation where we explicitly put in the
Coulomb interaction.  Accordingly the value of $\epsilon$ has been increased
from 3.7 MeV to 4.7 MeV.  There are now some changes from the results of
Figs. 1 and 2, but not a great deal.  Specially the deduced value of the slope
$\tau$ is again the lowest at $T=T_c$ and rises both below and above this
temperature. $T_c$ is $1.1275\epsilon$ and is the critical temperature in the
lattice gas model. The results for $\tau$ are summarised in Fig. 4 where it is
seen that the explicit inclusion of the Coulomb interaction has not modified
the predominant characteristics observed in calculations without explicit
inclusion of the Coulomb interaction.
Here $\tau$ was obtained from linear fits of fragmentation
distributions in  $\log Y(A)$ {\it vs } $\log A$ plots.
%
%%%%%% begin of replacement%%%%%%%%
That is, $\tau$ is determined by minimizing the $\chi^2$ defined as:
\begin{equation}
\chi^2 = \sum_i (F(A_i)-F_i)^2.
\end{equation}
Here $F(A_i)\equiv \log Y(A_i) = \mbox{const} -\tau \log A_i$
is the fitted yield for fragment of size $A_i$, and
$F_i\equiv\log Y_i$ is the corresponding simulated yield.
To maintain sufficient statistics and to exclude the largest cluster,
only fragments of sizes between 1 and 12 were used.
%%%%%%% end of replacement %%%%%%%%
%
In Fig. 5 for completeness we have
shown a $Y(Z)$ against $Z$ curve.  This is the type of curve that are typically
presented as experimental results.

Figs 6, 7 and 8 give our results for Au on Au.  We plot both $Y(A)$ against
$A$ and $Y(Z)$ against $Z$.  For molecular dynamics without explicit inclusion
of the Coulomb interaction we have used $\epsilon=3.7$ MeV and for calculation
with explicit inclusion of the Coulomb interaction we have used
$\epsilon=4.7$ MeV.  However now the results are very different for the two
cases.  In one (Fig. 6) there is a minimum in $\tau$ at $T=T_c$.  A second
spike at $T=0.4T_c$ is indicative of a percolating cluster.  In Fig. 7
with explicit inclusion of the Coulomb interaction, the percolating cluster has
disappeared.  We also see that below the critical temperature the
$\tau$ values from the two calculations begin to diverge.  With Coulomb
explicitly included the minimum in $\tau$ has disappeared and one can get
a value of $\tau$ much less than 2.  These results are summarised in Fig. 8.
Our results are in qualitative agreement with the experimental results
for Au on Au.  Experimental results are given as a function of the beam
energy and thus we need a conversion from temperature to beam energy.
This conversion needs to be done carefully because at initialisation, which is
the starting point of our calculation we can expect that some energy is
already in collective motion and does not appear as thermal excitation.  The
model does not include this aspect.  In Ref.\cite{pan1} a phenomenlogical
mapping of temperature to beam energy was deduced from \cite{li2}.
As an estimate only
if we assume that at the initial time $3/8$ of the initial energy is stored
in collective motion then a beam energy of 35MeV/nucleon would correspond to
$0.3T_c$.  Our calculated value of $\tau$ is then 1.4 compared to the
experimental value of 1.25 \cite{msu}.  As in experimental data the calculated
$Y(A)$ against $A$ deviates from a power law with higher excitation energy.
We nonetheless deduce a effective value of $\tau$ from a very approximate
fit and these are shown in Fig. 8.
We regard $\tau$ as a measure of global feature of $Y(A)$ , although
power-law fits  are poor at high temperatures.
To maintain sufficient statistics, we used fragments
of size between 1 and 20 for high temperatures ($T \geq T_c$) when heavier
fragments are rare. For low temperatures ($T < T_c$) larger
fragments were also included.
Our calculation at about $1.1T_c$ fits the data for beam energy 100
MeV/nucleon.
For the calculation with Coulomb interaction included we use $T_c$
merely as an energy scale; there is no implication that $T_c$ is the
critical temperature of the system.
The principal point we want to emphasize is that we have reproduced the most
significant features of the data for Au on Au as contrasted with those for
Ar on Sc, namely that in the former case there is no minimum in the value of
$\tau$ and that the value of $\tau$ can be significantly below 2.

\section{Conclusion}

This problem started out as an effort to understand why the fragmentation
data for Au on Au are so different from that of Ar on Sc and if
the data totally ruin all validity of the simple concepts used in the
lattice gas model.  The calculation done here suggests that the lattice gas
model is reasonable for medium mass collisions; it probably would have been
as valid for collisions of very large masses but for the very
large Coulomb force which begins to make its presence felt and destroys the
simple predictions.
It has often been assumed that the larger the system of colliding nuclei the
better is the chance of learning about phase transition in nuclear matter.
However larger colliding masses also bring in much larger Coulomb forces
and it will be necessary to take into account of the Coulomb effects before
the signals for phase transitions can be understood.  With large masses
the mean field of the protons are very different from that of the neutrons
and theories must be able to treat them differentially.  There clearly are
needs for simple theories which are able to handle this difference.

\section*{Acknowledgments}

We are very grateful to Miniball Group of Michian State University for
providing with recent experimental results for central Au on Au collisions.
This work was supported in part by the Natural
Sciences  and Engineering Research Council of Canada and
by the FCAR fund of the Qu\'ebec Government.

\newpage
\section*{Figure captions}
\begin{description}
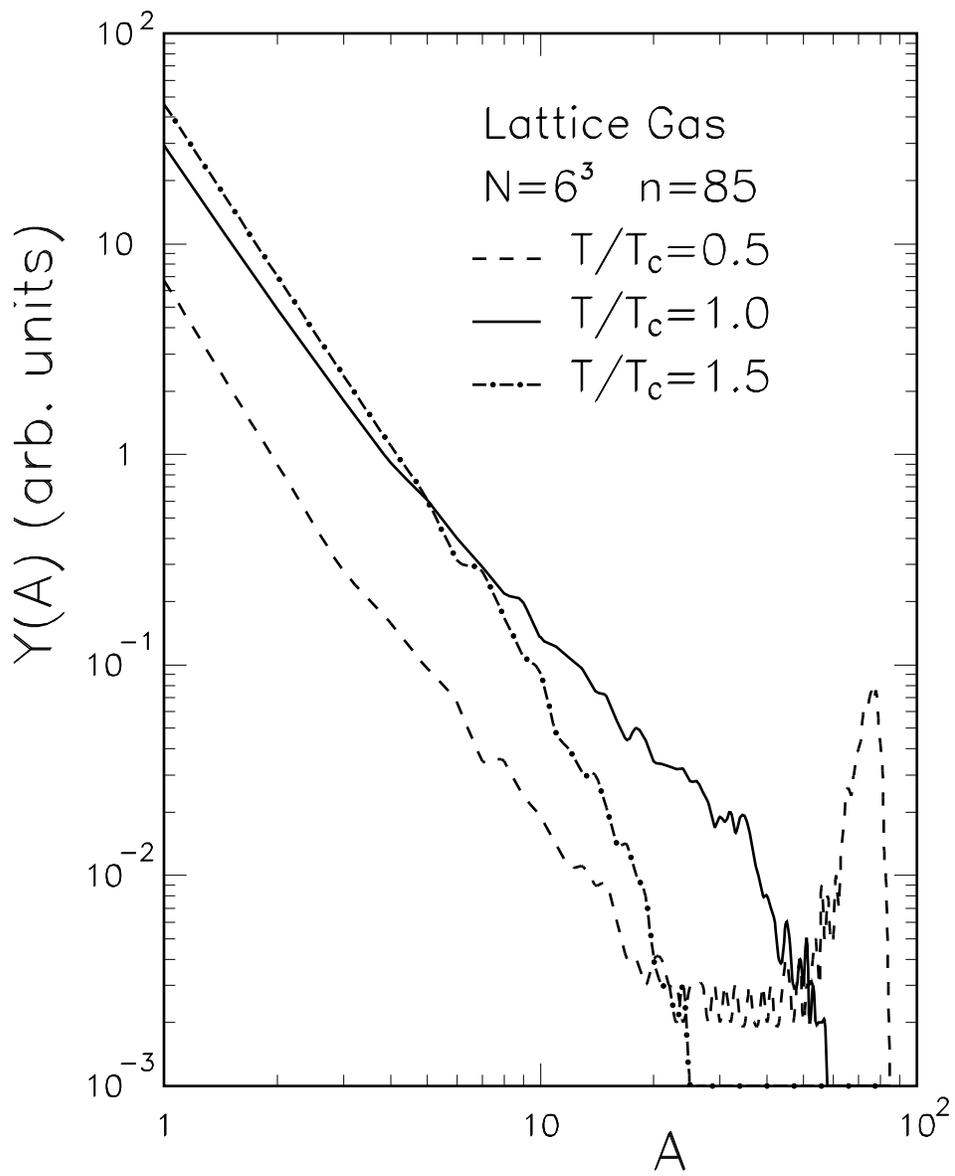
\item[Fig. 1] The mass yield distributions obtained from the lattice gas
        model for lattice $N=6^3$ and $n=85$, at temperatures
        $T/T_C=$0.5, 1.0 and 1.5. Here $T_C=1.1275\epsilon$ is
        the thermal critical temperature.
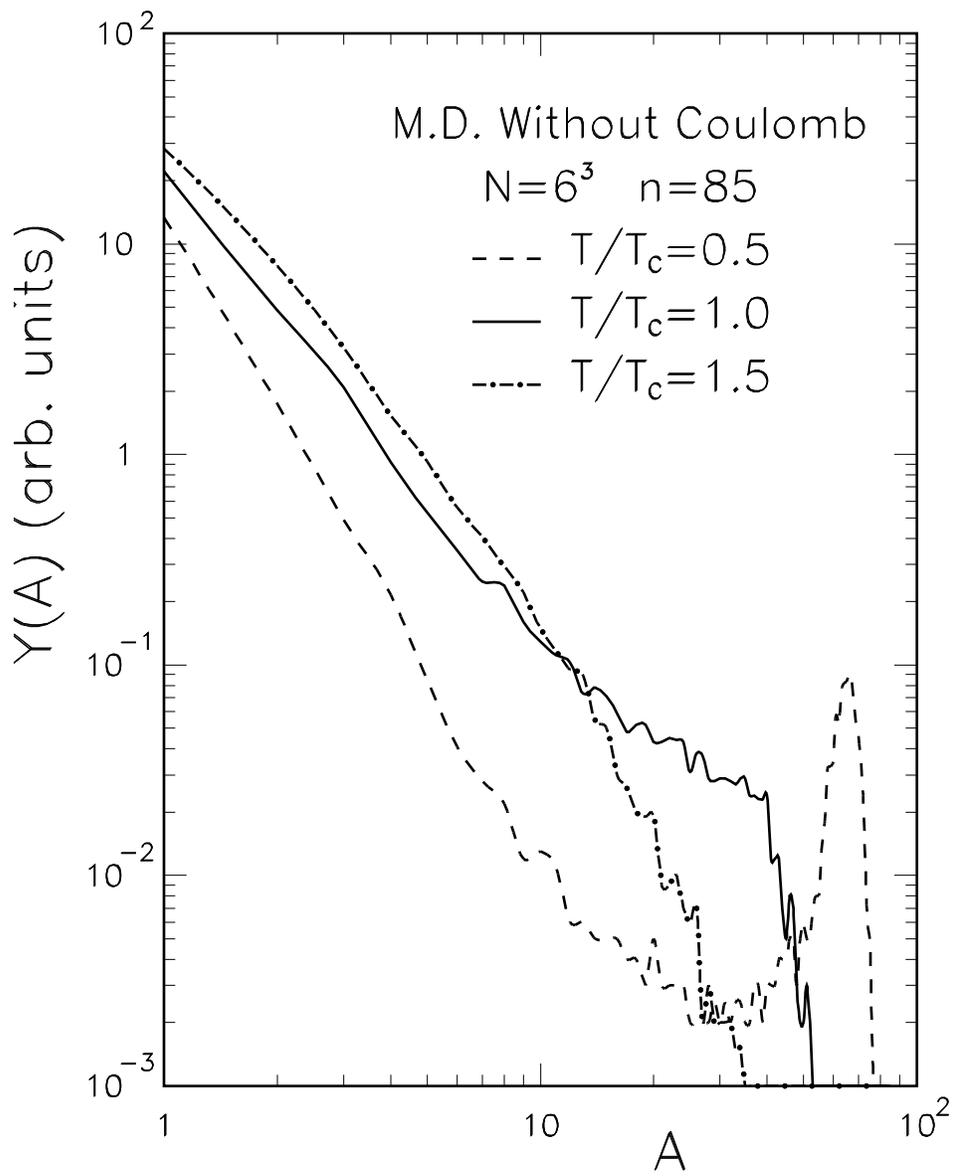
\item[Fig. 2] The mass yield distributions obtained from molecular
        dynamics calculations without the inclusion of Coulomb interaction.
	The lattice size, number of nucleons and temperatures are the
	same for the corresponding curves in Fig. 1.
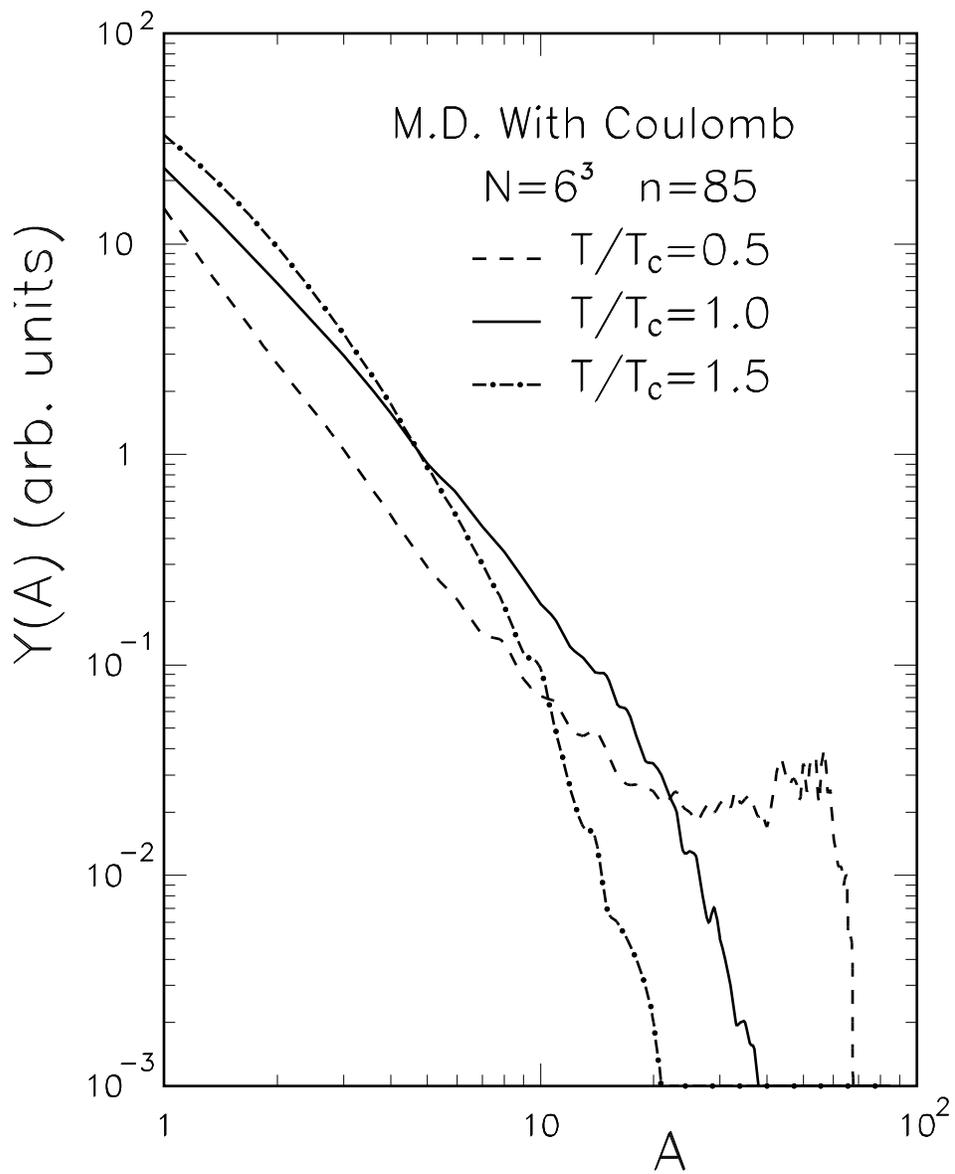
\item[Fig. 3] The same as Fig. 2, but the Coulomb interaction is taken
	into account.
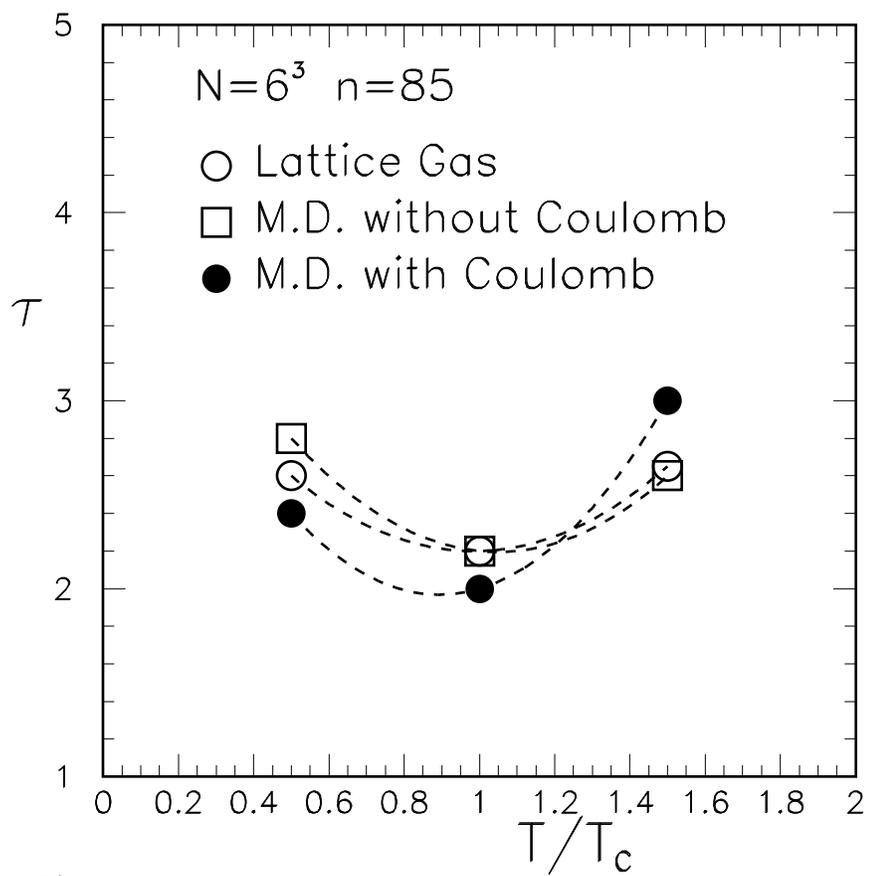
\item[Fig. 4] The value of $\tau$ obtained from lattice gas model and molecular
	dynamics calculations are plotted as a function of temperature
	for $N=6^3$ and $n=85$.
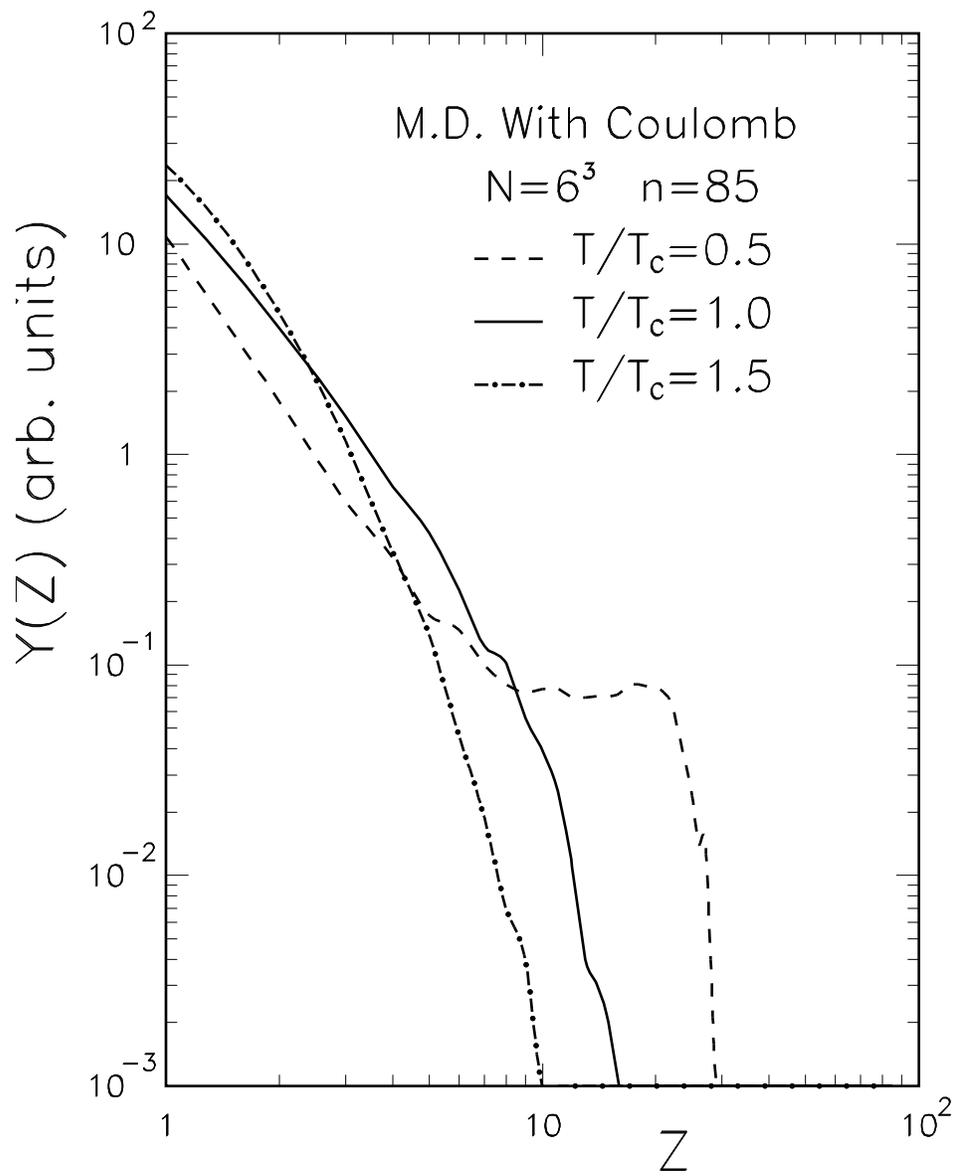
\item[Fig. 5] The charge yield distributions obtained from the molecular
	dynamics calculation with Coulomb interaction are shown.
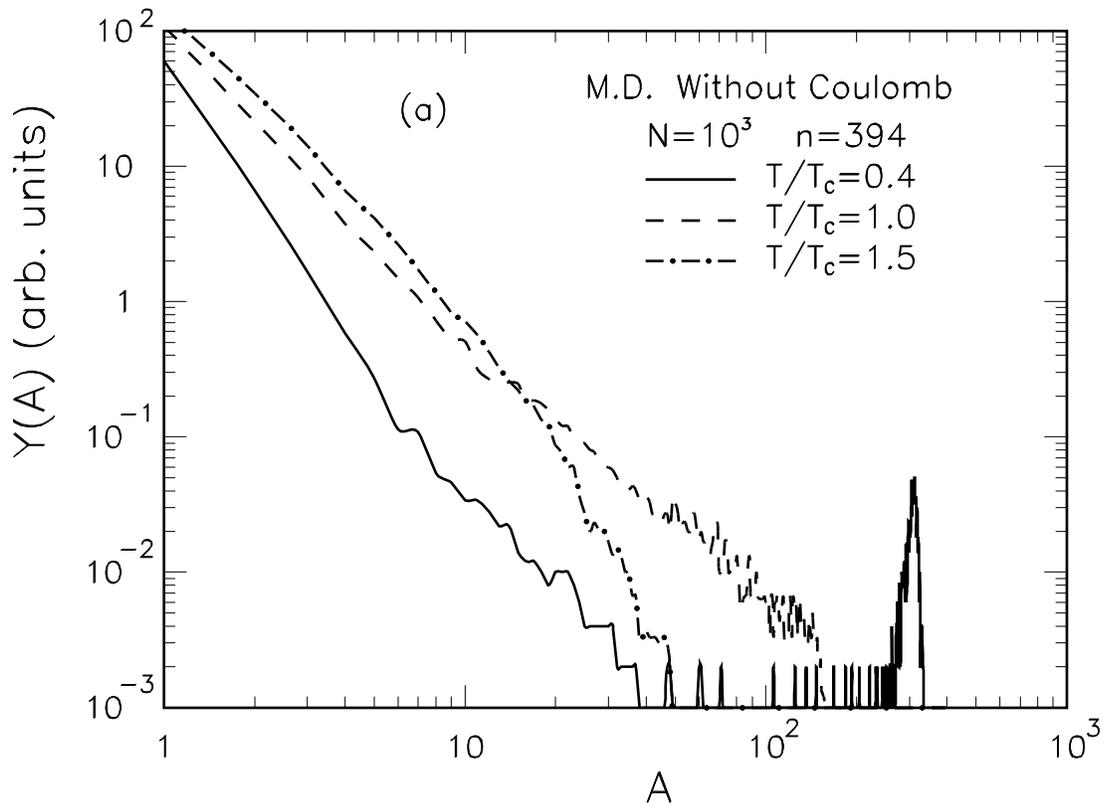
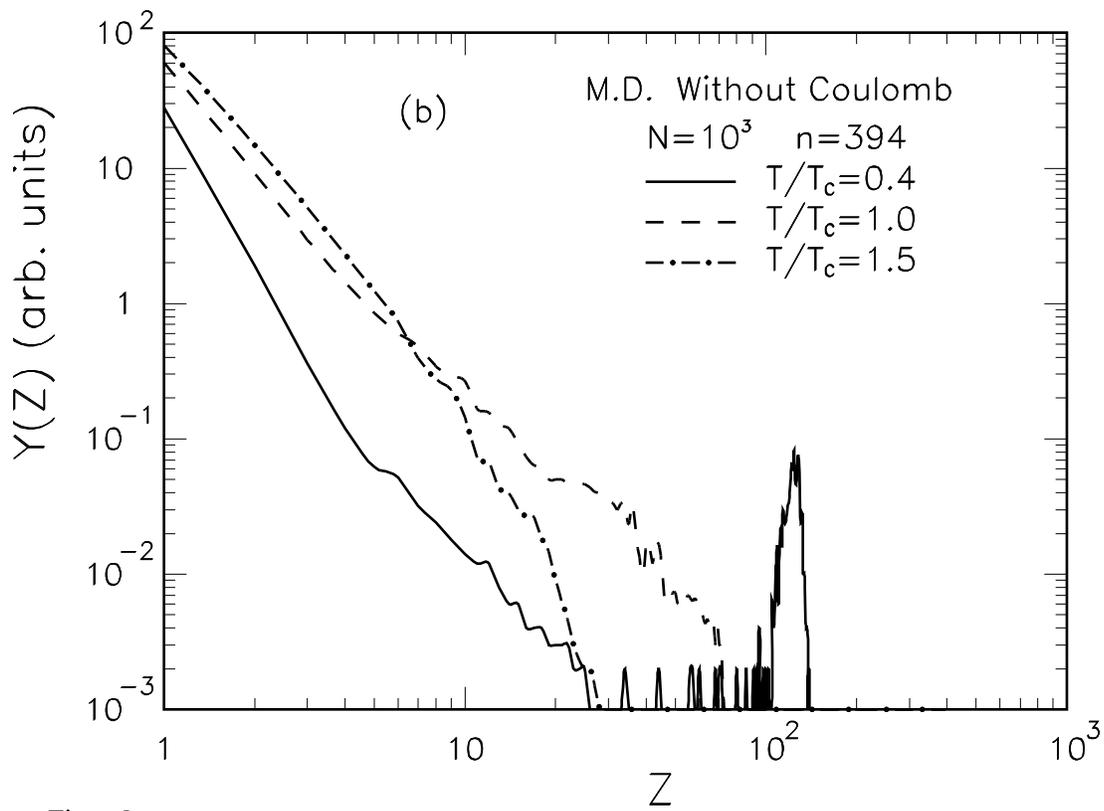
\item[Fig. 6] The mass and charge yield distributions for Au on Au
	collisions obtained from the molecular dynamics without Coulomb
	interaction.
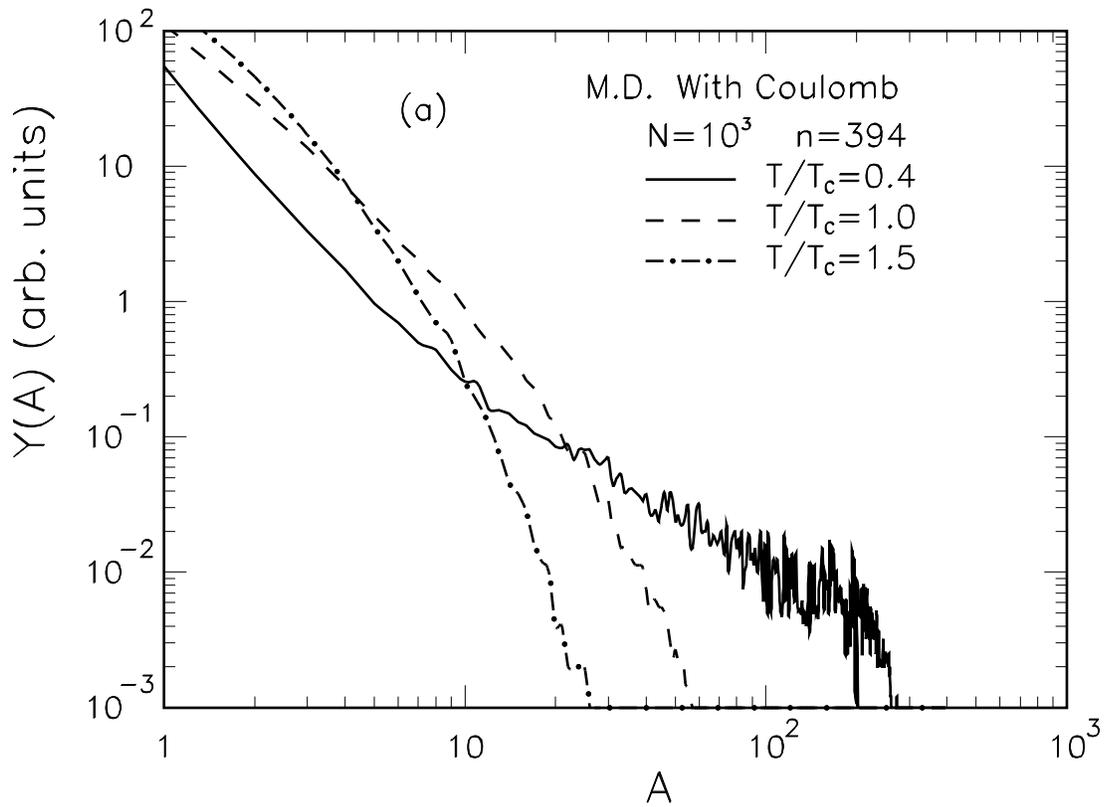
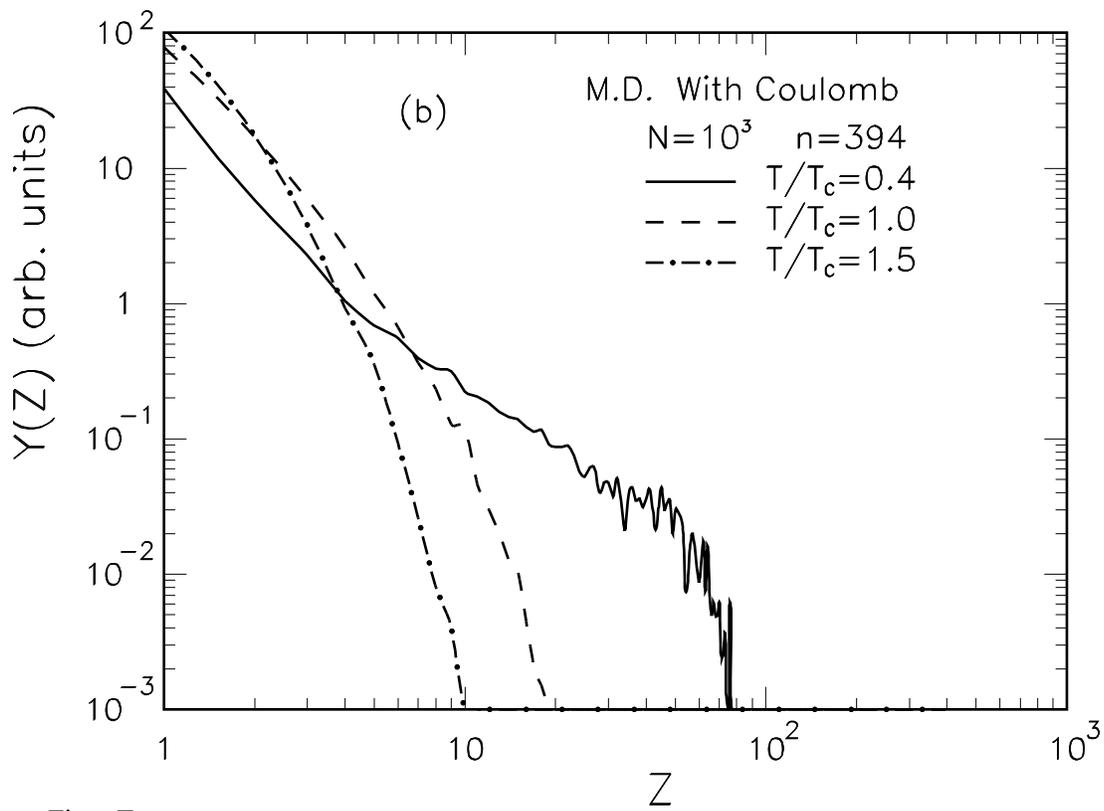
\item[Fig. 7] The same as Fig. 6, but with Coulomb interaction.
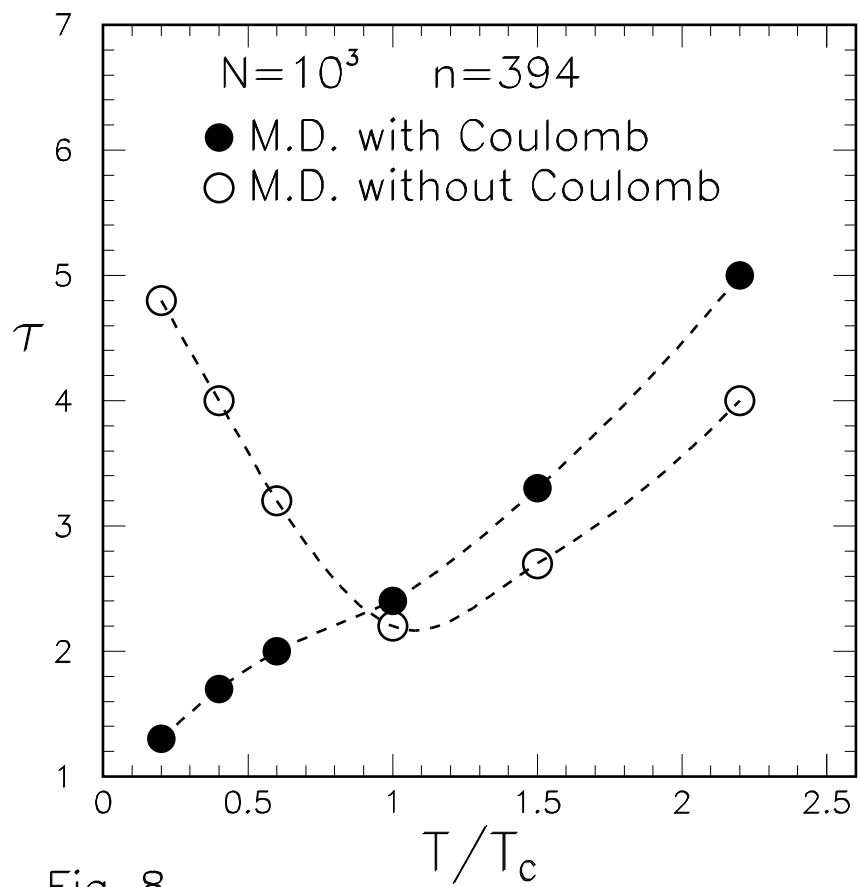
\item[Fig. 8] The value of $\tau$ in Au on Au collisions obtained from
molecular
	dynamics calculations with and without Coulomb interaction are plotted
	as a function of temperature.
\end{description}

\end{document}